\documentstyle[prb,aps,twocolumn,tighten,psfig,epsf,floats]{revtex}
\begin{document}
\draft
\title{{\em Ab initio} many-body calculations of static dipole polarizabilities
of linear carbon chains and chain-like boron clusters}
\author{Ayjamal Abdurahman}
\address{Technische Universit\"at Dresden,Institut f\"ur Physikalische
Chemie und Electrochemie,D-1062 Dresden, Germany}
\author{Alok Shukla}  
\address{Physics Department, Indian Institute of Technology, Bombay, 
Powai, Mumbai 400 076, India}
\author{Gotthard Seifert} 
\address{Technische Universit\"at Dresden,Institut f\"ur Physikalische
Chemie und Electrochemie,D-1062 Dresden, Germany}

\maketitle
\begin{abstract} 
In this paper we report a theoretical study of the static dipole
polarizability of two one-dimensional structures: (a) linear carbon 
chains C$_{n} (n=2-10)$ and (b) ladder-like planar boron chains 
B$_{n} (n=4-14)$. The
polarizabilities of these chains are calculated 
both at the Hartree-Fock and the correlated level by
applying accurate ab initio quantum chemical methods. Methods such as 
restricted Hartree-Fock, multi-configuration self-consistent field, 
multi-reference configuration-interaction method, 
M{\o}ller-Plesset second-order perturbation theory, and coupled-cluster
singles, doubles and triples level of theory were employed. Results
obtained from ab initio wave-function-based methods are compared with
the ones obtained from the density-functional theory. 
For the clusters studied, directionally
averaged polarizability per atom for both the systems is seen to increase with
the chain size.
\end{abstract}

%\pacs{}

\section{Introduction}
The role that atomic and molecular clusters will play in future
nanotechnologies is indisputable.~\cite{bonin,dres} The experimental
progress in this field has been breathtaking, and novel
applications have been found in areas such as molecular transport and 
optoelectronics.~\cite{bonin,dres,bera} However, theoretical
research in this area can also play a very important role in that, 
by undertaking calculations on clusters of different types, 
it can help the experimentalists in identifying novel structures for
investigation. {\em Ab initio} calculations on structural and electronic    
properties of atomic clusters are frequently performed, and the results are 
put to test in the experiments.~\cite{metclus} One such property of clusters
is their electric-dipole polarizability whose experimental and theoretical
determination is an area of intense research.~\cite{bonin}
The measurements of dipole polarizability are frequently used by the 
experimentalists to 
 characterize the nature of atomic and molecular species.~\cite{bonin}
It describes the response of the electron cloud of the given molecular
system to the presence of
a d.c. electric  field, and thus is easily amenable to experiments. Since the
static polarizabilities are the zero-frequency limits of the corresponding
dynamic quantities, they also provide information about the response
of the system to off-resonant a.c. fields. 
Most of the theoretical calculations of both the structural and electronic
properties such as static polarizabilities of atomic clusters are 
performed within the framework of the density functional theory (DFT). 
Despite the fact that DFT has enjoyed
indisputable success in solid state physics 
and quantum chemistry as a computationally cheap routine tool 
for large-scale investigations, it has the drawback that results depend 
highly on the chosen functional, and cannot be improved in a systematic way.
Wave-function--based {\em ab initio } quantum-chemical  techniques on the
other hand are free from this flaw, and provide a large array of 
methods of different accuracy and computational cost. 
Moreover,
the prediction of reliable values of dipole polarizabilities and 
hyperpolarizabilities by rigorous quantum chemical methods has made 
significant contributions, and added new vigor, to  the search of novel
optical materials~\cite{bera,metclus}. 
Thus, 
in order to obtain reliable estimates for dipole polarizabilities, and also
to cross check the DFT-based results, it
is worthwhile to investigate the electron correlation  
effects in a systematic way by using the quantum-chemical many-body 
techniques. 
In this work, we present fully size-consistent {\em ab initio}
calculations to the static dipole polarizability of linear
carbon clusters C$_{n} (n=2-10)$ and chain-like
boron clusters B$_{2n} (n=2-7)$, of increasing size. The reason behind
our focus on one-dimensional structures is that quantum confinement
due to reduced dimensions, combined with the possible delocalization of
the electrons along the backbone, can lead to enhanced linear and nonlinear
susceptibilities of these structures, as compared to their three-dimensional
counterparts.  In the present study, the electron-correlation effects have 
been taken into account by various size-consistent
methods: multi-reference configuration interaction, second-order
M{\o}ller-Plesset perturbation theory, coupled-cluster singles and
doubles (CCSD), and coupled-cluster singles and doubles with
the perturbative treatment of the triples (CCSD(T)). All earlier calculations
on these systems, with the exception of B$_4$, were performed within the 
framework of DFT, with which we compare our results. 
Next, we briefly review
the state-of-the-art of research on these two types of clusters.

Carbon clusters have been the subject of research
for decades as possible key materials for future
nanotechnologies~\cite{dres}. For the smaller systems, up to and
including those containing nine atoms, linear neutral, positively and
negatively charged clusters are generated and detected in
experiments~\cite{yang}. The structures and energetics  of linear
carbon clusters are well studied by employing coupled-cluster
approaches.~\cite{ragha,watt}  Recently, we studied
the ground state of an infinite carbon chain at the {\em ab initio} level
using various  many-body
approaches including CCSD(T).~\cite{ayjamal} Lou et al. have investigated the
influence of an electric field 
on the energetic stability of linear carbon chains~\cite{lou}. Recently
Fuentealba~\cite{fuentealba1} has calculated the static dipole
polarizabilities of carbon chain using density functionals of the
hybrid type in combination with the finite-field method. He showed that dipole
polarizabilities are an important quantity for the identification of
clusters with different numbers of atoms and even for the separation
of isomers. Here we compare our many-body-methods based polarizabilities with
those computed by Fuentealba~\cite{fuentealba1} using the DFT approaches.

Boron is a trivalent
element with the valence shell configuration  $s^2p^1$. Although
compared to carbon, the valence shell of boron is electron deficient,
it still exhibits $sp^2$ hybridization with strong directional chemical
bonds.~\cite{adams} In composite materials, for small content, boron
tends to form a linear chain, while as its content increases it can
form structures ranging from two- to three-dimensional.~\cite{adams} 
As far as isolated clusters containing boron are concerned, experimentally,
they have been studied by Anderson and 
coworkers,~\cite{anderson} and by La Placa et al.~\cite{placa}
In addition, Chopra et al.~\cite{chopra} and Lee et al.~\cite{lee} have
experimentally synthesized the boron-nitrogen nanotubes and cage-like
boron-nitrogen structures. Several authors have also reported theoretical 
calculations on the boron clusters.~\cite{ray,kawai,koutecky,ricca,bostan,%
bostan0,bostan00,bostan1,bostan2,bostan3,bostan4,bostan5,reis,reis1}
Boustani et al.~\cite{bostan} have shown theoretically that, similar to carbon,
boron has a strong potential to form
stable nanotubular structures.
Boustani and coworkers recently studied small
cationic~\cite{bostan0} and neutral boron clusters~\cite{bostan00}
and obtained structures that are fundamentally different from crystal
subunits of the well-known $\alpha-$ and $\beta-$ rhombohedral phases
of boron, which consist mainly of $B_{12}$ icosahedra. They classified
the boron clusters into four topological groups: convex and 
spherical~\cite{bostan1} , quasiplanar~\cite{bostan2} and 
nanotubular~\cite{bostan3}.
The quasiplanar and convex structures can be considered as fragments
of planar surfaces, and as segments of hollow spheres,
respectively. The main focus of their theoretical work has been to
ascertain the structures of larger boron clusters in terms of
a small number of building blocks.~\cite{bostan4,bostan5}
However, recently Sabra and Boustani~\cite{bostan5} studied the
ground-state energetics of ladder-like quasi-one-dimensional clusters of
boron by quantum chemical methods. They concluded that such 
structures are not the lowest in energy. However, because of the proximity of
their energy to that of the true ground-state geometries, they can be regarded
as metastable states.~\cite{bostan5} 
Thus, with some experimental manipulation, it may be possible to realize 
such structures in laboratory. Keeping this possibility in mind, we decided
to compute the static polarizabilities of ladder-like quasi-one-dimensional
structures of boron by {\em ab initio} many-body methods. In addition to
the quantum-chemical calculations, we also perform the DFT-based calculations 
of static dipole polarizabilities of these clusters using the same basis set,
so as to understand the influence of electron-correlation effects on the
polarizabilities of these systems.
Recently, Reis and coworkers  computed the static dipole 
polarizabilities of rhombic B$_4$ using various quantum-chemical 
methods,~\cite{reis} and several other boron clusters B$_{n} (n=3-8,10)$ 
within the framework of DFT, employing a variety of exchange-correlation 
functionals.~\cite{reis1}
However, unlike the quasi-one-dimensional geometries considered by us,
Reis et al.~\cite{reis1} performed these calculations on the
ground-state geometries of Boron clusters optimized earlier by 
Boustani.~\cite{bostan4} We compare our many-body static polarizabilities
of various boron clusters to those reported by Reis et al.~\cite{reis,reis1}   
in order to understand the influence of the cluster structures on their static 
polarizabilities.

The remainder of the paper is organized as follows. In section \ref{method}, 
the applied methods and computational details are briefly described. 
The results are 
then presented and discussed in section \ref{results}. Finally, our
conclusions are presented in  
section \ref{conclusions}. 

\section{Methods and computational details}
\label{method}
For the closed-shell clusters, first the polarizabilities are calculated 
by using the restricted Hartree-Fock (RHF) method, and thereafter the 
electron-correlation effects are included via the
M{\o}ller-Plesset second-order perturbation theory (MP2), and the 
coupled-cluster (CC)
techniques. For the open-shell clusters, the calculations are initiated by
the multi-reference self-consistent-field (MCSCF) method, while the 
electron-correlation effects are taken into account by the multi-reference
configuration-interaction (MRCI) method.
In order to calculate static dipole polarizability first we
performed calculations without an external electric field, and then we added 
an external electric field of strength $0.001$
a.u. along the $x$, $y$ and $z$ axis separately. Stability of the results with
respect to the value of the field was carefully examined by performing some
calculations for various other values of the field strength. However, when 
we perform high-level correlated calculations,  the expectation value of the
dipole moment is not directly available. Therefore, to calculate the
static dipole polarizabilities,  
we have adopted a finite-difference formula in which the diagonal
polarizability tensor elements are obtained through the second derivative
of the total energy with respect to the external electric
field. The field-dependent total energy is 
used in the following finite-difference formula:
\begin{center}
$$\alpha_{jj}=-\left[
\frac{\partial^{2}E(\varepsilon_{j})}
{\partial{\varepsilon_{j}^{2}}} \right]_{\vec{\varepsilon}=0}
=-\lim_{\varepsilon_{j} \to 0}\frac{E(\varepsilon_{j})+E(-\varepsilon_{j})-2E(0)}{\varepsilon_{j}^{2}}=\lim_{\varepsilon_{j} \to 0}2\frac{E(0)-E(\varepsilon_{j})}{\varepsilon_{j}^{2}}$$
\end{center}
where $E(\varepsilon_{j})$ is total energy with respect to field
$\varepsilon_{j}=0.001$ a.u. and $E(0)$ is total energy without
field. This equation holds only for centrosymmetric systems.\\ 
For the linear carbon chain, assuming the $z$ axis is
the chain direction, we calculated the parallel ($\alpha_{zz}$) and
perpendicular ($\alpha_{xx}$) components of the static dipole
polarizability. Our calculations are performed using the geometry
reported by Watts et al.~\cite{watt}.  
Since the ground state of even number of carbon atoms is triplet and
of odd number of carbon atoms is singlet we calculated the static
dipole  polarizabilities of even 
number of carbon atoms i.e., $C_{4}$, $C_{6}$, $C_{8}$ and $C_{10}$
for its ground state by the MCSCF and the MRCI methods, whereas for odd
number of carbon atoms i.e., $C_{3}$, $C_{5}$, $C_{7}$ and $C_{9}$ we
calculated them by using the RHF, MP2, CCSD, and
the CCSD(T) methods.  
All calculations were performed with the MOLPRO molecular orbital {\em
ab initio} program package~\cite{molpro} by employing Sadlej basis
sets~\cite{sadlej} which was specially constructed for the calculation
of dipole polarizabilities.\\
For the chain-like boron clusters we assumed that the boron atoms were lying 
in the $xy$ plane, with the chain direction along the $x$-axis. We first 
optimized the
geometry of each cluster, i.e., $B_{4}$, $B_{6}$, $B_{8}$, $B_{10}$,
$B_{12}$ and $B_{14}$ for its ground state at the B3LYP/6-31+G(d) level of
approximation by using the GAUSSIAN98 program~\cite{gauss}. The
ground state is singlet for $B_{4}$, $B_{10}$, $B_{12}$
$B_{14}$ and triplet for $B_{6}$, $B_{8}$. Then we
calculated the parallel  ($\alpha_{xx}$), transverse  ($\alpha_{yy}$)
and perpendicular  ($\alpha_{zz}$) components of the 
static dipole polarizabilities with standard polarized
valence double-zeta (VDZ) basis sets at the Hartree-Fock
and correlated level e.g., MRCI, MP2, CCSD and CCSD(T) by
employing MOLPRO molecular orbital {\em ab initio} program
package~\cite{molpro}. Although the basis set which we used in chain-like
boron clusters is a rather small basis set, a larger set would have been
computationally too expensive when we prolong the chain. It has been
shown further in  Ref.~\cite{reis1} that using the larger triple-zeta
basis set aug-cc-pVTZ does not have a large effect on the calculated
polarizabilities for the cluster $B_{4}$. Therefore, for these clusters the
chosen basis set  should be sufficient.
\section{Results and discussions}
\label{results}
\subsection{Linear carbon clusters}
\begin{figure}
\caption{The static dipole polarizability per atom of linear carbon 
clusters. The line has been plotted to guide the eyes.}
\centerline{\psfig{file=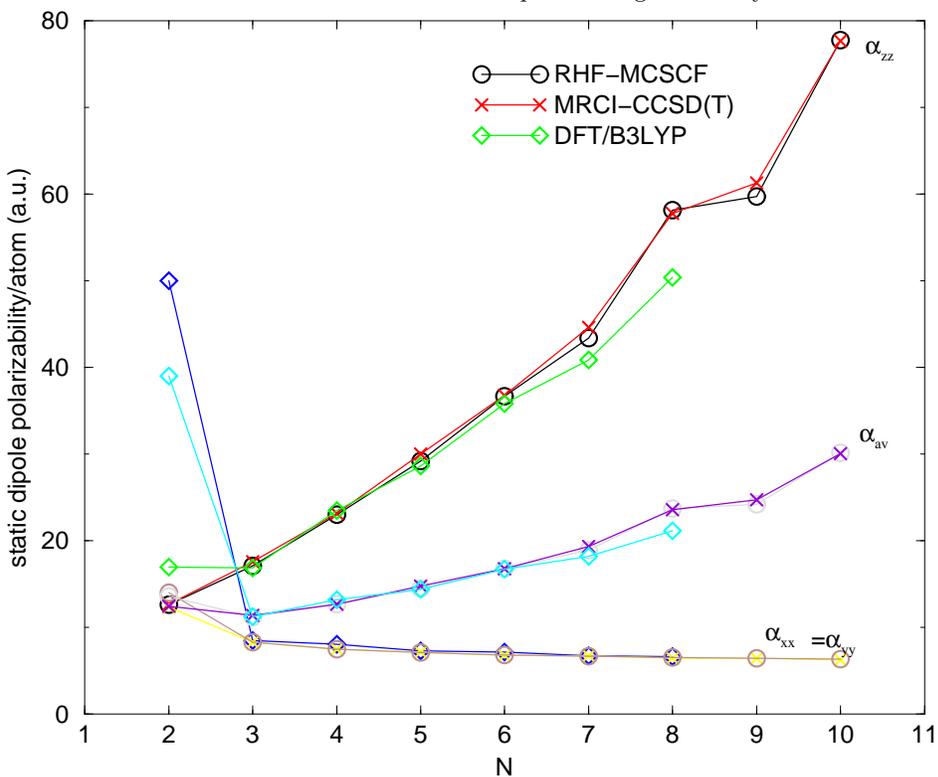,width=12.5cm,angle=-90}}
\label{fig-polcar}
\end{figure}

 The calculated Cartesian components of
 static dipole polarizabilities and average polarizabilities
 $\alpha_{av}=(\alpha_{zz}+2\alpha_{xx})/3$ for linear carbon clusters
 are presented in Table \ref{tab-polcar}. Figure \ref{fig-polcar} presents our
calculated polarizabilities per atom, plotted as a function of the number
of atoms in the chain ($N$). Additionally, for the sake of comparison, in the 
same figure we have also plotted the DFT-based results of 
Fuentealba~\cite{fuentealba1}. It is not possible for us to compare our
 results to experiments because of the absence of any polarizability data
on the carbon chains. From Fig. \ref{fig-polcar}  it is obvious that 
parallel component of the 
static dipole polarizability per 
atom, $\alpha_{zz}/N$, increases roughly linearly from $C_{2}$ to
 $C_{10}$, whereas  perpendicular components $\alpha_{xx}/N$ and 
 $\alpha_{yy}/N$ are essentially constant as a function of $N$. As far as
the comparison of our results with the DFT results of 
Fuentealba~\cite{fuentealba1} is concerned, the agreement is generally 
very good on all components of polarizabilities except for the case of
$N=2$. For $N=2$, however, Fuentealba~\cite{fuentealba1} reports an
 anomalously large value of $\alpha_{xx}$ (and hence $\alpha_{xx}/N$),
making it in disagreement with his values of $\alpha_{xx}/N$ computed for 
higher values of $N$.
From Fig. \ref{fig-polcar} it is also clear that  the 
directionally-averaged polarizability per atom $\alpha_{av}/N$,
 also  shows an overall
 increase as a function of the chain length. 
Since polarizability is an
extensive quantity, therefore, for very large number of atoms in the
chain ($N \rightarrow \infty$), $\alpha_{av}/N$ should approach its bulk value.
However, from our results it is obvious that for $N=10$, $\alpha_{av}/N$ is
still increasing as a function of $N$, exclusively because of the increase in 
the parallel component  $\alpha_{xx}/N$.
One can understand the increase in $\alpha_{xx}/N$ as a function of $N$ 
on the intuitive grounds based upon the behavior of $\pi$ electrons. $\pi$
 electrons (of which the carbon chain has two per atom),
as against the $\sigma$ electrons, are highly delocalized along
 the chain direction. Therefore, it will take much larger cluster sizes before
their response to an external field approaches that in the bulk.

The other somewhat surprising aspect of our results for the carbon chains
is the generally excellent agreement obtained between the DFT values, and 
the many-body
values of the static polarizabilities. This means that for the static
polarizabilities of carbon chains, DFT is able to describe the electron
correlation effects quite well. It is also rather interesting to note 
that MP$2$ method provides a
 theoretical description of these clusters quite close to that obtained
 with the CCSD(T) method. A similar effect was observed by 
Maroulis~\cite{mor} in the polarizability calculations of a system 
composed of two water molecules.

\subsection{Chain-like boron clusters}
\begin{figure}
\centerline{\psfig{file=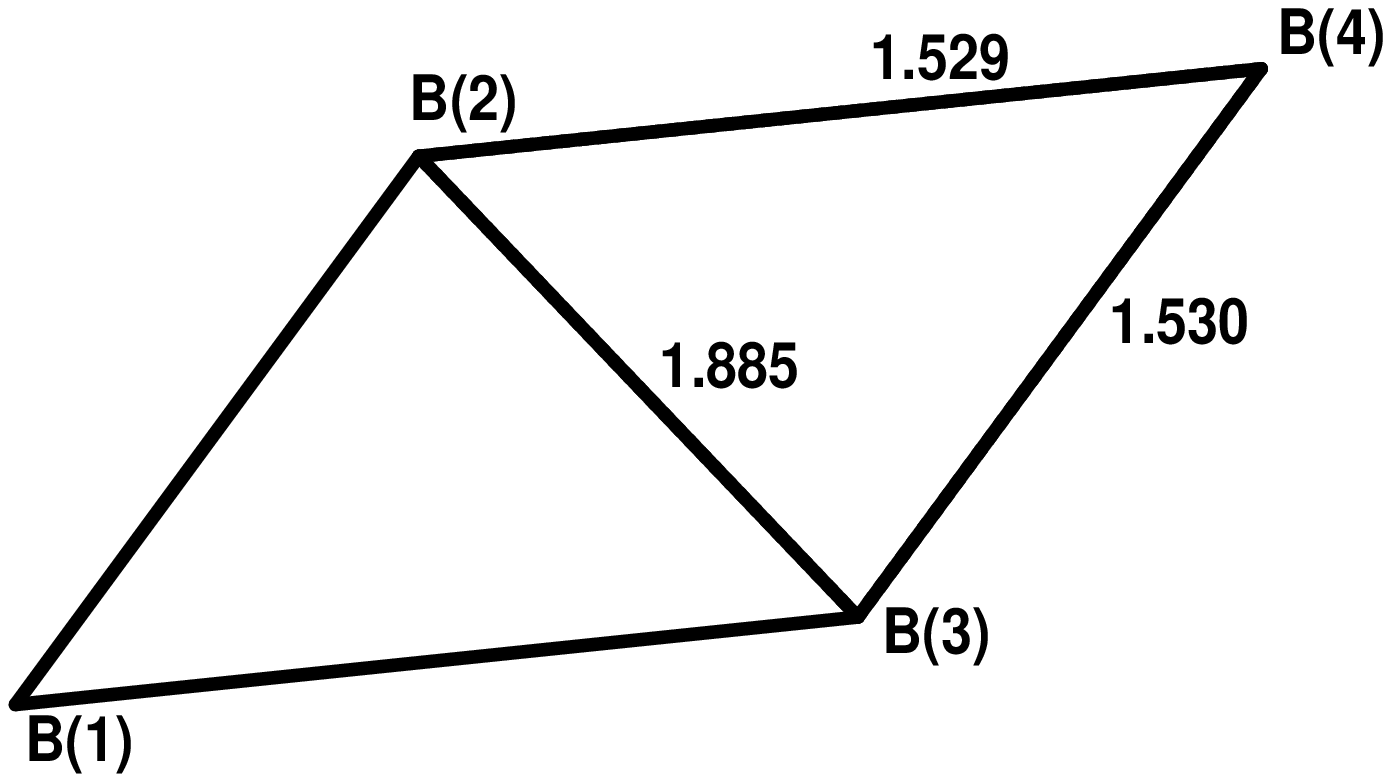,width=6.5cm,angle=90}\psfig{file=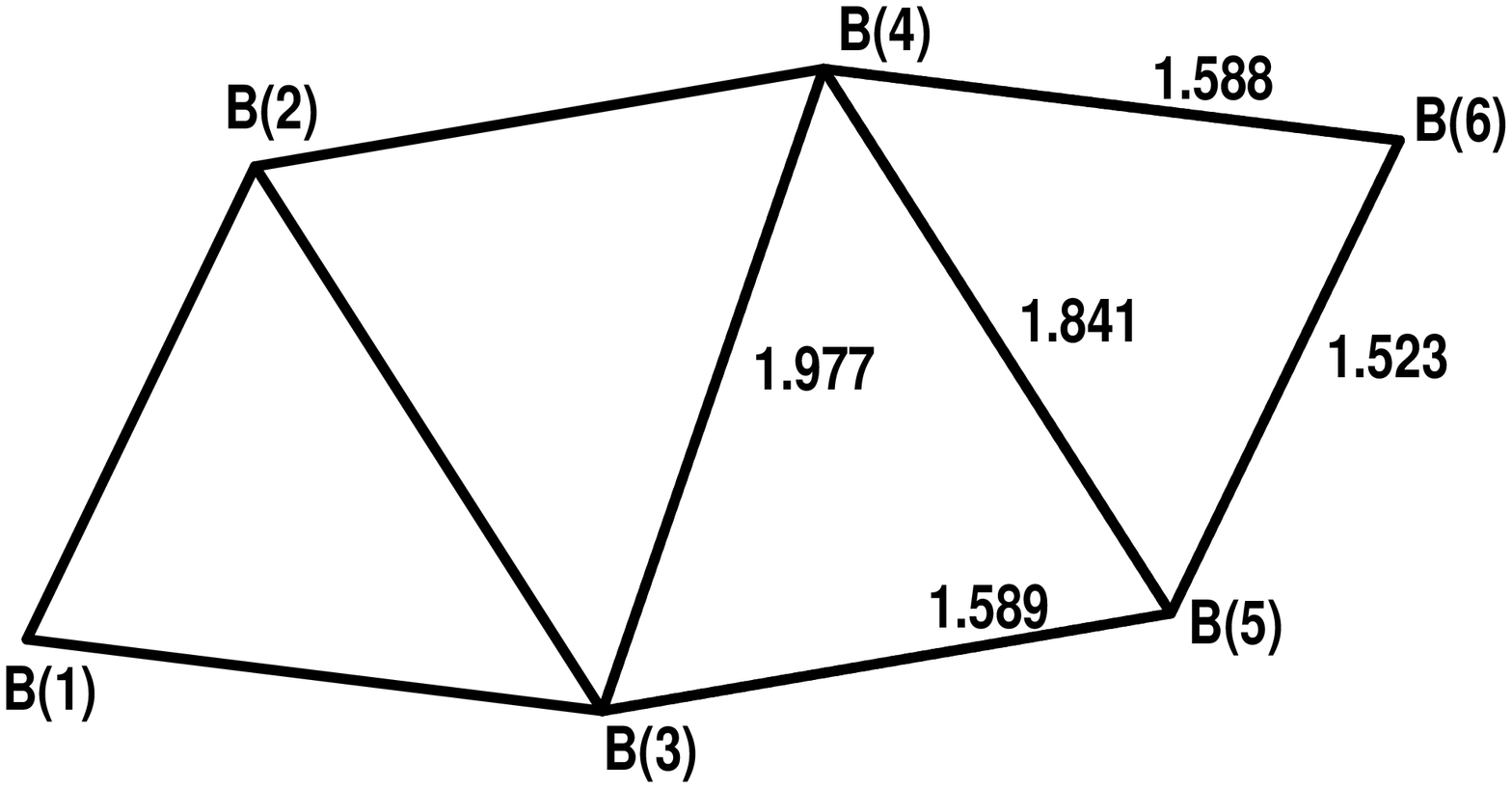,width=6.5cm,angle=90}}
\centerline{\psfig{file=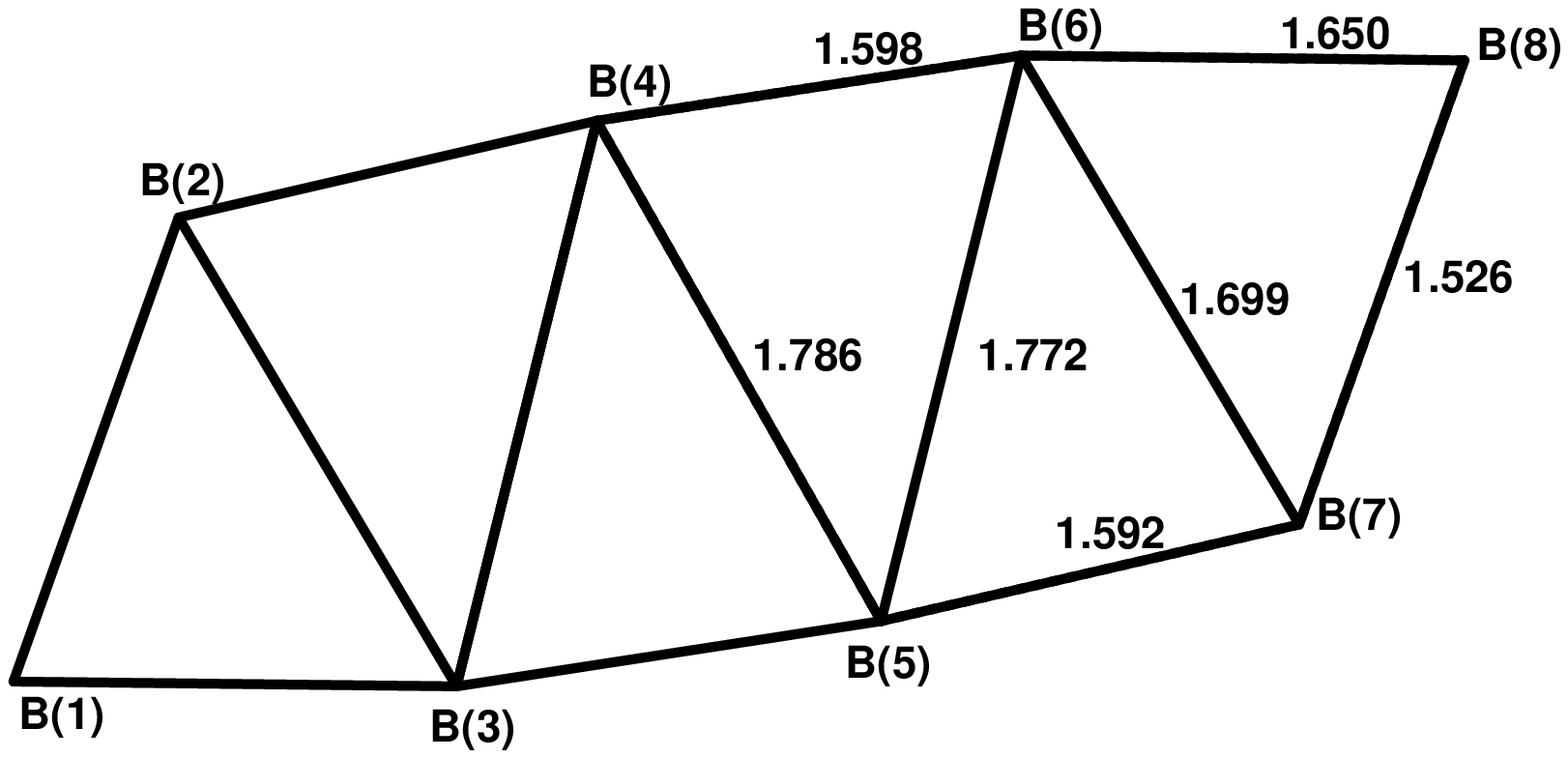,width=9.5cm,angle=90}}
\centerline{\psfig{file=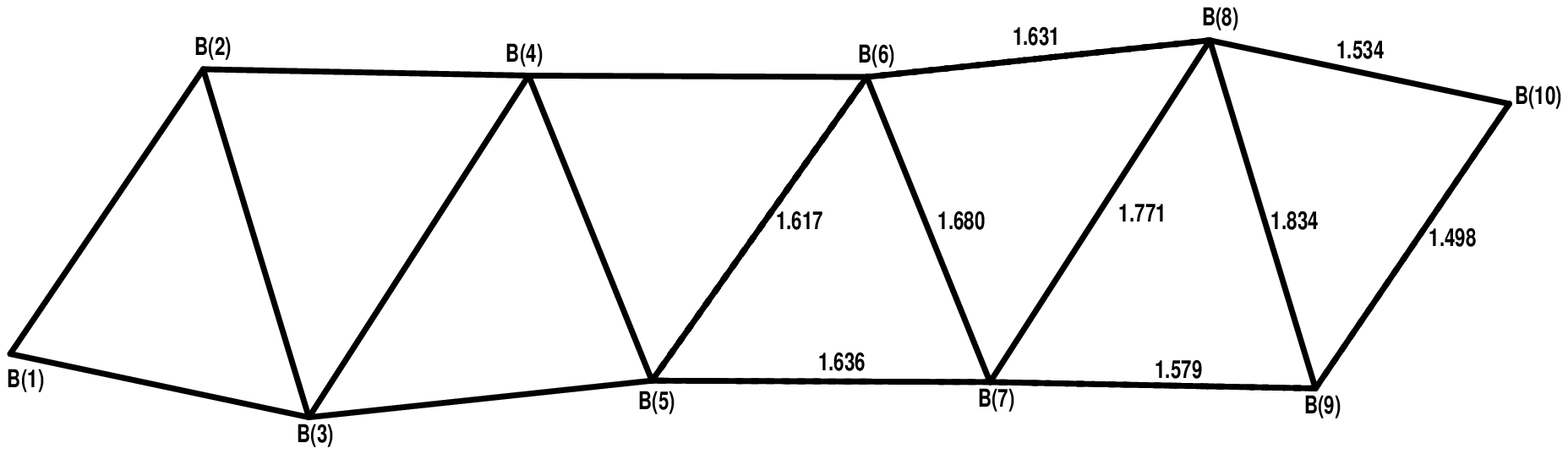,width=17.5cm,angle=90}}
\centerline{\psfig{file=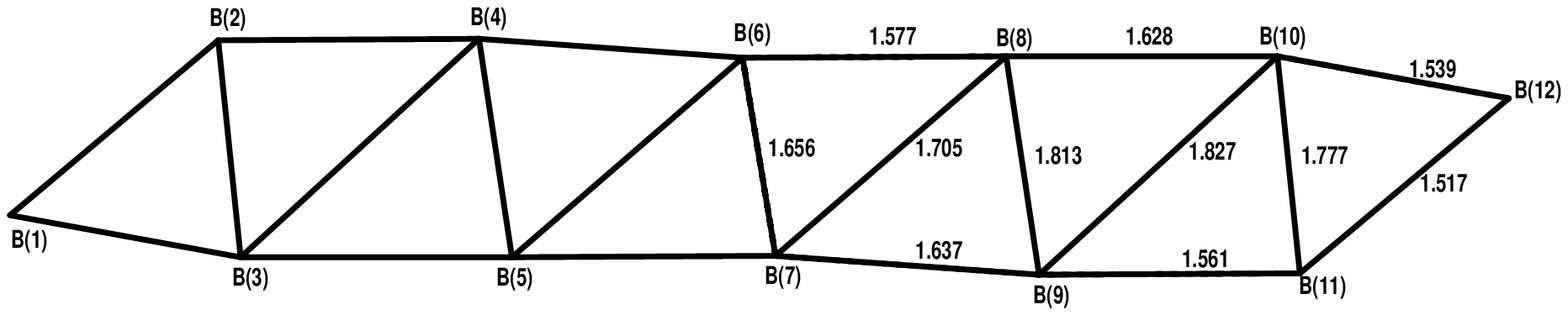,width=18.5cm,angle=90}}
\centerline{\psfig{file=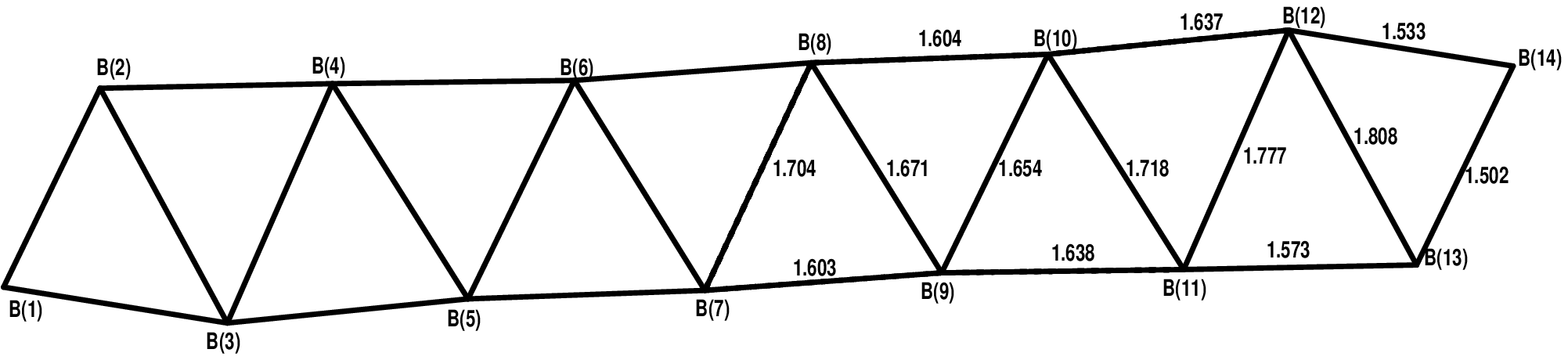,width=19.5cm,angle=90}}
\caption{The chain-like structure of boron clusters.}
\label{fig-borclus}
\end{figure}

Earlier Sabra et al.~\cite{bostan5} had shown that strictly one-dimensional 
chains of boron are unstable. They demonstrated that boron prefers to
form a zig-zag ladder-like quasi-one-dimensional structure.~\cite{bostan5}
Therefore, in the present work we have concentrated on the identical
structures of boron B$_{2n}(n=2-7)$, which, as shown in Fig. \ref{fig-borclus},
can be obtained by adding boron dimers to B$_{4}$, which has a parallelogram
structure. First we optimized the ground-state geometry of each of these 
clusters by employing B3LYP/6-31+G(d) method, and these optimized geometries
are given in Fig. \ref{fig-borclus}. From optimization we found that
the system is completely centrosymmetric. The optimized geometries of each
cluster are comparable to those obtained by Boustani~\cite{bostan4} who 
optimized the structures of elemental, convex and quasiplanar boron
clusters B$_{n}(n=2-14)$ at the RHF level with the $3-21G$ basis set.  

\begin{figure}
\caption{The static dipole polarizability per atom of chain-like boron 
clusters. The line has been plotted to guide the eyes.}
\centerline{\psfig{file=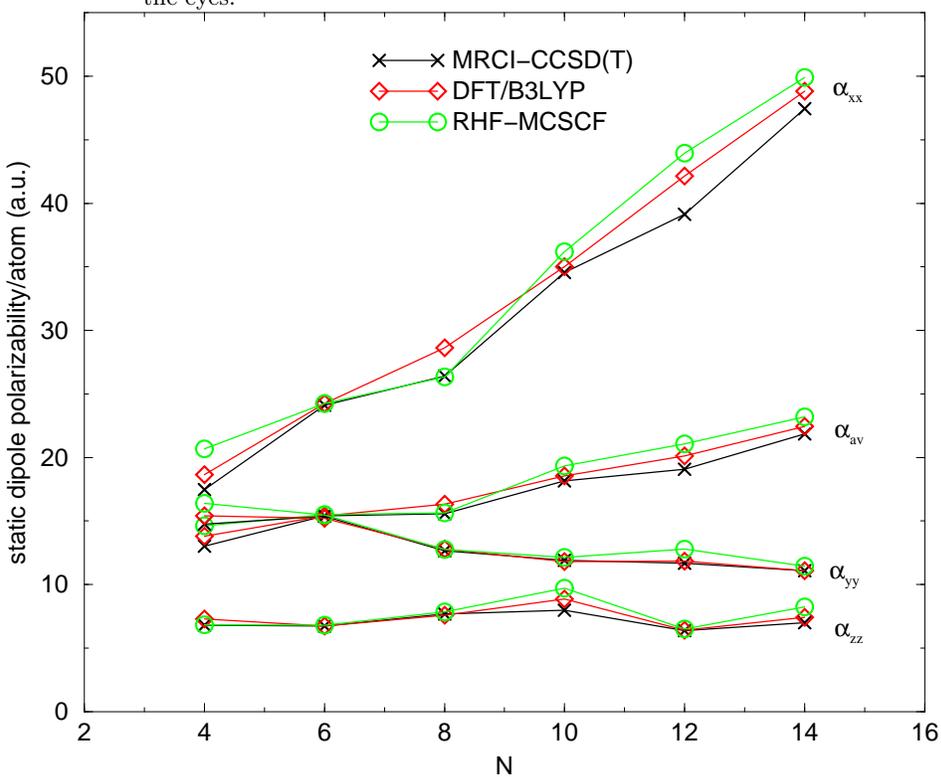,width=12.5cm,angle=-90}}
\label{fig-polbor}
\end{figure}

Results of our calculation on longitudinal $(\alpha_{xx})$, transverse
$(\alpha_{yy})$, perpendicular $(\alpha_{zz})$, and directionally-averaged 
polarizabilities
$\alpha_{av}=(\alpha_{xx}+\alpha_{yy}+\alpha_{zz})/3$ at the HF and
correlated level, as well as at the DFT level are presented in 
table \ref{tab-polbor}. The polarizabilities per atom based upon this data
are plotted as a function of the number of atoms in the cluster in 
Fig. \ref{fig-polbor}. An inspection of the table and the figure reveals the 
following
trends: (i) The longitudinal static dipole polarizability per atom
$\alpha_{xx}/N$ increases almost linearly with $N$,  the component  
$\alpha_{yy}/N$ shows a gradual decrease, while $\alpha_{zz}/N$ exhibits
saturation. (ii) For the triplet ground state clusters B$_6$ and B$_8$,
the polarizabilities computed by the MCSCF 
and the MRCI methods are in almost complete agreement
indicating that the MCSCF
method has already captured the most important correlation effects.
(iii) For the remaining clusters whose ground state is singlet, the inclusion
of electron correlation effects leads to a reduction of the $\alpha_{xx}$
component, while the other components are rather unaffected. For example,
the CCSD(T) value of $\alpha_{av}$ for B$_{14}$ is about 6 \% smaller compared
to its RHF value. (iv) Similar to the case of carbon chains, for all boron
clusters considered here, generally there
is very good agreement between the polarizabilities computed by the best
wave-function methods (MRCI and CCSD(T)), and the ones computed using the
DFT/B3LYP approach. 
Thus, in this case also, the DFT is
able to account for the electron correlation effects quite well.

 Since there are no experimental results for the static 
dipole polarizabilities for the
ladder-like structures of boron, we compare our 
results to the theoretical results obtained by other 
authors.~\cite{reis,reis1} 
First considering the case of rhombic B$_4$, for the average
static dipole polarizability $\alpha_{av}$, we obtained 51.99 a.u. for
 with the CCSD(T) approach while Reis et al.~\cite{reis}
report a CCSD(T) value 60.00 a.u. for the same quantity. The $\alpha_{xx}$ and
$\alpha_{yy}$ values from our calculations
cannot be directly compared to those reported by Reis et al.~\cite{reis} 
because of the different orientations of the $x$ and the $y$ axes in their
calculations. However,  for $\alpha_{zz}$, whose values can be compared 
directly, Reis et al.~\cite{reis} report the value 39.5 a.u., while we 
obtained 27.12 a.u. for the same quantity. 
Although the optimized geometries, as well as the basis set used
by Reis et al.~\cite{reis} were somewhat different from ours, we still believe
that those factors cannot explain the difference of $\approx$ 12 a.u. in the
values of $\alpha_{zz}$. 
However, clearly it is this disagreement---the reasons behind which are not
clear to us---which is primarily responsible for
the disagreement in the values of $\alpha_{av}$ observed between our results
and those of Reis et al.~\cite{reis}

Besides B$_4$, there are no theoretical results on the larger ladder-like
clusters of boron. However, in another paper Reis et al.~\cite{reis1} 
reported DFT based calculations
of the static polarizabilities of convex and quasiplanar B$_n$
($n=3-10$) clusters whose geometries were optimized earlier by
Boustani.~\cite{bostan4} Therefore, in order to understand the effect of the 
geometric structure on the polarizabilities of boron clusters, we compare 
our results on B$_6$, B$_8$, and B$_{10}$, with the corresponding isomers
studied by Reis et al.~\cite{reis1} For B$_6$ Reis et al.~\cite{reis1} 
considered a benzene-like hexagonal geometry
with the $D_{2h}$ symmetry, and computed the value $\alpha_{av}=101.3$ a.u.
For B$_8$ also Reis et al.~\cite{reis1} considered a ring-like structure
with the $D_{7h}$ symmetry and reported $\alpha_{av}=114.6$ a.u. Finally,
for B$_{10}$ they considered a quasiplanar structure with the 
$C_{2h}$ symmetry
and calculated $\alpha_{av}=143.7$ a.u.~\cite{reis1}
These can be compared with our DFT values of $\alpha_{av}$ of the 
ladder-like B$_6$, B$_8$, and B$_{10}$ which were obtained to be 92.35 a.u., 
130.65 a.u., and 185.53 a.u., respectively. From the comparison it is clear
that although the polarizability of the benzene-like B$_6$ is larger than
that of the ladder-like B$_6$, however, for clusters containing larger
number of atoms (B$_8$ and B$_{10}$) quasi-one-dimensional ladder-like
structures are more polarizable than the planar structures. Although, no
theoretical results on $\alpha_{av}$ for the planar structures of B$_{12}$ and
B$_{14}$ are available, however, it is clear that even for those clusters
ladder-like structures will be obtained to be more polarizable. The fact that
for larger number of atoms, ladder-like boron chains will be more polarizable
than quasiplanar isomeric structures of boron, can be understood based upon
intuitive arguments. As Sabra et al.~\cite{bostan5} showed by explicit 
calculations, with the increasing size, the $\pi$-electron population
of these ladder-like chains increases. Since the $\pi$-electrons are quite 
delocalized along the chain direction, their response to the electric fields
directed along the chain direction will be quite large leading to large values
of $\alpha_{xx}$ obtained in our calculations. The fact that $\alpha_{xx}$ 
increases quite rapidly with the increasing number of atoms also
confirms this hypothesis. Although, the quasiplanar structure of boron also
have $\pi$ electrons due to the sp$^2$ hybridization, however, their response
to the external field is distributed in two directions due to their
two-dimensional character, leading to smaller polarizabilities.
Therefore, it is the combined effect of $\pi$ electrons and the reduced
dimensionality which makes the ladder-like chains of boron more polarizable
than its quasiplanar counterparts.
\section{Conclusions and Future Directions}
\label{conclusions}
In conclusion, we have reported a systematic {\em ab initio} study of the 
static dipole polarizability of the linear carbon chains and the 
ladder-like boron chains employing Hartree-Fock, many-body, and the DFT-based
approaches. For closed-shell clusters the polarizabilities computed by the
RHF method were generally within 5\%---6\% agreement with the ones computed
by the CCSD(T) method. Similarly for the open-shell clusters the MCSCF
polarizabilities were found to be in very good agreement with the ones
computed by the MRCI method. Additionally, DFT-based results on the
polarizabilities were found to be in very good agreement with the ones
computed by the many-body methods. This suggests the possibility that by
employing computationally less expensive approaches such as RHF, MCSCF, DFT 
etc.  one can perform similar calculations on even larger and more complex 
clusters, and obtain reasonable results on static polarizability. We believe
that such a line of investigation should be pursued in future calculations. 

 Our results demonstrate that in both the systems the component of the
polarizability along the chain direction, as well as 
the average polarizability, increase with the chain
size. This is fully consistent with presence of the delocalized
$\pi$-electrons in these one-dimensional clusters. Thus both types of clusters
should be useful in nonlinear optical applications as well. 
Of late, the molecular
transport properties of carbon chains have been of much interest to 
physicists,~\cite{c-transport} because of the presence of the delocalized
$\pi$-electrons in them. 
However, our polarizability calculations point to the presence of the 
delocalized $\pi$-electrons also in the ladder-like boron clusters, thus
rendering them
possibly useful in molecular-transport-based applications. First principles
studies of the excited states, and the transport properties of such clusters 
will be the subject of future investigations.

\acknowledgements
One of us (A.A.) is grateful to Professor T. Wolff and Graduiertenkolleg
Struktur-Eigenschafts-Beziehungen bei Heterocyclen for financial support.

\clearpage
\newpage
\clearpage
\newpage
\clearpage
\newpage
\begin{table}
\caption{Linear carbon clusters: Static dipole polarizabilities (in a.u.) calculated with the Sadlej basis set.}
\begin{tabular}{llllll}
\hline
Atoms & Methods   & $\alpha_{xx}=\alpha_{yy}$ & $\alpha_{zz}$    &
$\alpha_{av}$ \\
\hline
C$_2$           &  MCSCF     & 28.00    & 25.22     & 27.07   \\
                &  MRCI      &24.57     &25.24      &24.79 \\
             &DFT/B3LYP$^a$& 100.00   &33.90   & 78.00  \\
\hline
C$_3$               & RHF & 24.72     & 51.16     & 33.53   \\
                    & MP2 & 24.46     & 51.68     & 33.53   \\
                    & CCSD(T) & 24.82     &52.66     & 34.10  \\
                    &DFT/B3LYP$^a$&25.30    &50.50   & 33.70   \\
\hline
C$_4$       & MCSCF      &29.86  &91.88 &50.54    \\
            & MRCI      &29.71  &92.38 & 50.60   \\
           &DFT/B3LYP$^a$& 32.10   &93.70   & 52.60    \\
\hline
C$_5$              & RHF & 35.42    & 145.74     & 72.19   \\
                    & MP2 & 35.38     &149.74      &73.50     \\
                    & CCSD(T) & 35.50     &149.88      & 73.63     \\
                   &DFT/B3LYP$^a$&36.20    & 142.90  & 71.70    \\ 
\hline
C$_6$       & MCSCF      &40.81  &219.88 & 100.50     \\
            & MRCI      &40.45  & 220.40&100.44    \\
            &DFT/B3LYP$^a$& 42.80   & 214.70  & 100.00     \\
\hline
C$_7$               & RHF & 46.54    & 303.50    &132.19   \\
                    & MP2 &46.22      &314.80      &135.75      \\
                    & CCSD(T) & 46.74     & 312.30     & 135.26   \\
                  &DFT/B3LYP$^a$&47.00    &285.80   & 127.00     \\ 
\hline
C$_8$       & MCSCF      &51.82  &464.78 &189.47     \\
            & MRCI      &51.34  & 462.11& 188.26    \\
             &DFT/B3LYP$^a$& 52.60   & 403.00  & 169.00      \\ 
 \hline     
C$_9$               & RHF & 57.58    & 537.12    & 217.43   \\
                    & MP2 & 57.20     & 555.04    &223.15      \\
                    & CCSD(T) & 57.64     & 551.54     & 222.27   \\
\hline    
C$_{10}$ & MCSCF      & 62.85  & 777.51& 301.07     \\
       & MRCI      & 62.29  & 776.39& 300.32   \\
\end{tabular}
$^a$ Taken from Ref.~\cite{fuentealba1}. \\
\label{tab-polcar}
\end{table}
\begin{table}
\caption{Chain-like boron clusters: Static dipole polarizabilities (in a.u.)
calculated with the VDZ basis set.}
\begin{tabular}{llllll}
\hline
Atoms & Methods   & $\alpha_{xx}$ & $\alpha_{yy}$ & $\alpha_{zz}$    & $\alpha_{av}$ \\
\hline
B$_4$               & RHF &82.72    &65.54     &27.30 & 58.52\\
                    & MP2 &66.34    &57.28    &27.30   &50.31\\
                    & CCSD(T) &69.88     &58.98     &27.12  &51.99  \\
                    &DFT/B3LYP&74.60   &61.60  &29.20   &55.13 \\
\hline
B$_6$              & MCSCF &145.45     &92.86     &40.78  &93.03 \\
                    & MRCI &144.64      &92.16  &40.28   &92.36 \\
                    &DFT/B3LYP&145.51   &91.15  &40.39     &92.35  \\
\hline
B$_8$               & MCSCF &210.72    &101.93    &62.73  &125.12 \\
                    & MRCI &211.28    &101.17   &61.54     &124.66\\
                    &DFT/B3LYP&229.05  &102.40    &60.51      &130.65\\
\hline
B$_{10}$               & RHF &361.86    &121.24    &97.00    &193.37\\
                    & MP2 &362.08    &120.86    &63.54   &182.16\\
                    & CCSD(T) &345.70    &118.90    &79.82  &181.47 \\
                    &DFT/B3LYP&350.20    &118.00  &88.40    &185.53 \\
\hline
B$_{12}$               & RHF &527.32    &153.60   &77.90    &252.94\\
                    & MP2 &445.82    &129.52      &76.04  &217.12\\
                    & CCSD(T) &469.70     &140.06     &76.58  &228.78\\
                    &DFT/B3LYP&505.80   &142.20  &76.80    & 241.60\\
\hline
B$_{14}$               & RHF &698.90    &160.16     &115.36  &324.81\\
                    & MP2 &680.52    &152.68        &75.92     &303.04\\
                    & CCSD(T) &664.48     &155.08    &97.66    &305.74 \\
                    &DFT/B3LYP&683.80    &155.20    &104.00  &314.33   \\
\end{tabular}
\label{tab-polbor}
\end{table}

\end{document}